\documentclass{article}
\usepackage{spconf,amsmath,graphicx}
\usepackage{amsfonts}
\usepackage{amssymb}
\usepackage{algorithm}
\usepackage[noend]{algpseudocode}
\usepackage{multirow}
\usepackage{multicol}
\usepackage{booktabs}
\usepackage{tabularx}
\usepackage{enumitem}
\usepackage[dvipsnames]{xcolor}
\usepackage{float}
\usepackage{todonotes}
\usepackage{url}
\usepackage{hyperref}

\usepackage{booktabs}

\usepackage{todonotes}

\makeatletter
\newcommand*\iftodonotes{\if@todonotes@disabled\expandafter\@secondoftwo\else\expandafter\@firstoftwo\fi}  
\makeatother

\newcommand*{\genbf}[1]{\ifmmode\mathbf{#1}\else\textbf{#1}\fi}

\title{Efficient Speech Representation Learning with Low-Bit Quantization}
\name{Ching-Feng Yeh, Wei-Ning Hsu, Paden Tomasello, Abdelrahman Mohamed}

\address{
Meta AI \\
\texttt{\{cfyeh,wnhsu,padentomasello,abdo\}@meta.com}
}

\begin{document}

%
\maketitle
\begin{abstract}
With the development of hardware for machine learning, newer models often come at the cost of both increased sizes and computational complexity. In effort to improve the efficiency for these models, we apply and investigate recent quantization techniques \cite{bit,sqwq} on speech representation learning models \cite{hubert}. The quantization techniques were evaluated on the SUPERB \cite{superb} benchmark. On the ASR task, with aggressive quantization to 1 bit, we achieved 86.32\% storage reduction $(184.42 \rightarrow 25.23)$, 88\% estimated runtime reduction $(1.00 \rightarrow 0.12)$ with increased word error rate $(7.06 \rightarrow 15.96)$. In comparison with DistillHuBERT \cite{distillhubert} which also aims for model compression, the 2-bit configuration yielded slightly smaller storage $(35.84 \leftrightarrow 46.98)$, better word error rate $(12.68 \leftrightarrow 13.37)$ and more efficient estimated runtime $(0.15 \leftrightarrow 0.73)$.
\end{abstract}


\section{Introduction}
\label{sec:introduction}

Modern machine learning technology has pushed the limits of related applications above and beyond in daily lives. As the performances improves, the number of parameters and the computational complexity of the models are also growing significantly \cite{attention, gpt, bert, roberta}. The growth in resource consumption not only means higher energy usage but also makes these applications less accessible. On the other hand, with the development of mobile and wearable devices, machine learning applications have been transitioning closer to the device side over the past few years. Given the growth in complexity of models and the need from edge devices, improving model efficiency has gained heavy interests and has bee widely studied \cite{bit, sqwq, xornet, distillhubert, quant_speed, quant_noise}. 

Among the numerous directions for improving model efficiency, quantization is particularly appealing, as quantization aims to keep the original model architecture but replaces the parameters with lower-precision alternatives for both storage saving and computation reduction. In addition, quantization typically casts parameters to lower-precision data types such as integers, which are more favorable on edge devices since integer operations are typically much cheaper and faster for the processors on these devices \cite{quant_speed,quarl}. However, quantization in nature converts numbers from continuous domains to discrete domains, therefore introduces quantization errors in computation and typically cause the model performance to degrade. Therefore, in the field of quantization-related research, minimal performance loss and maximal efficiency gain, or a better trade-off, has always been the pursuit. 

Recently, speech representation learning has been gaining popularity due to the high potential in unifying and generalizing the common components across different speech tasks such as automatic speech recognition (ASR) and keyword spotting (KS). Traditionally, the models for individual tasks are designed and trained independently from each other. While this practice works well for individual tasks, there exists a major redundancy between models for these tasks since many components serve similar purposes. For example, both ASR and KS models have modules converting speech signals to higher-level embeddings. Having a shared module instead of two separate ones for both tasks will minimize the redundancy. In that spirit, speech representation learning aims to train a unified model to generate embeddings from speech signals to be adopted by downstream tasks and therefore reduces the overhead introduced by individual tasks.

In this work, we investigated two recently proposed quantization techniques: 1) robustly binarized Transformer \cite{bit} 2) squashed weight quantization \cite{sqwq}. We analyzed these quantization techniques on top of the HuBERT \cite{hubert} model for speech representation learning and evaluated on the SUPERB \cite{superb} benchmark. From the experimental results, significant storage reduction $(184.42 \rightarrow 25.23)$ and estimated runtime improvement $(1.00 \rightarrow 0.12)$ were observed from applying an extreme 1-bit quantization (binarization) with a word error rate degradation $(7.06 \rightarrow 15.96)$ on the ASR task. Although the degradation is non-trivial, compared with recent compression approaches such as DistillHuBERT \cite{distillhubert}, quantization still offers a better trade-off between resource consumption and model performance.

\section{Efficient Low-Bit Quantization}

Quantization converts tensors from high-precision domains (typically floating-point numbers) to low-precision (typically integers or binaries) domains. While quantization provides benefits such as reductions both on storage and computation, it also presents challenges to be applied with minimal performance degradation compared with the original high-precision models. In this section, we summarize the techniques adopted in this work to produce a good trade-off between efficiency and performance for quantized models.

\subsection{Quantization Aware Training (QAT) and Straight-Through Estimator (STE)}

Among the wide variety of quantization strategies, quantization aware training (QAT) \cite{qat,qkd} has been popular for closing the performance gap between original and quantized models. Different from post-training quantization, QAT incorporates quantization operations during both the inference and gradient computation during training. This enables the model parameters to simulate quantization effects along with the training data so that they are more robust to quantization errors in later stages \cite{quant_noise}. 

A major challenge for QAT is how to propagate the gradients and update the model parameters with quantization in place. As quantization is in nature a clipping operation, theoretically the gradients are either zeros or indifferentiables at most operating points, meaning the model parameters won't be effectively updated. To resolve this, straight-through estimator (STE) \cite{ste} was proposed as an estimation for gradient updates. In STE, the forward pass still utilizes the model parameters in discrete domain, but in the backward pass the gating from quantization is bypassed in the chain rule and the gradients are directly applied onto the model parameters, as in equations \eqref{eq:ste_forward} and \eqref{eq:ste_backward}. From equation \eqref{eq:ste_backward}, the gradients are simply passed to the unquantized parameters as an approximation. In this work, both QAT and STE were adopted in model training. 

\begin{align}
    forward: \boldsymbol{y}&=Q(\mathbf{W})*\boldsymbol{x} + \mathbf{b}. \label{eq:ste_forward}
\end{align}
\begin{align}
    backward: \frac{\partial \boldsymbol{y}}{\partial \mathbf{W}} = \frac{\partial \boldsymbol{y}}{\partial Q(\mathbf{W})} * \frac{\partial Q(\mathbf{W})}{\partial \mathbf{W}} \stackrel{\footnotesize \mbox{STE}}{\approx} \frac{\partial \boldsymbol{y}}{\partial Q(\mathbf{W})}. \label{eq:ste_backward}
\end{align}

\subsection{Robustly Binarized Transformer (BiT)}
\label{subsec:binarized_transformer}

Conventionally, for {\it n}-bit quantization, real-valued numbers are converted into discrete counterparts, such as $\{0, 1, ..., 2^{n}-1\}$ for asymmetric cases and $\{-2^{n-1}, 1, ..., 2^{n-1}-1\}$ for symmetric cases with optionally a scaling factor $\alpha$. Recently, new quantization techniques has emerged beyond this simple formulation \cite{bit, sqwq, quant_noise}. Among the techniques, the robustly binarized transformer (BiT) \cite{bit} demonstrates smaller quantization errors and more aggressive yet more efficient model inference by applying quantization not only on parameters but also on activations.

The core idea of BiT is the two-set elastic quantization, where different formulations are applied to different numerical ranges. For example, the outputs from softmax operations will be positive only, while the weight in linear operations can be either positive or negative. This is referred to as the "two-set" quantization scenario, in which one is for asymmetric (positive only) and the other is for symmetric (both positive and negative). This enables better utilization for the precious bits in quantization. Given a scaling factor $\alpha \in R_{+}$ and a threshold $\beta \in R$, a tensor $\mathbf{X}$ can be quantized as in equation \eqref{eq:elastic_quantization}. Both $\alpha$ and $\beta$ can be stored as additional model parameters and updated during QAT with gradients through STE.

\begin{align}
    \mathbf{X}_{Q}  = \begin{cases}
      \alpha * Clip(\frac{\mathbf{X} - \beta}{\alpha}, 0, 1), & \text{if } \mathbf{X} \in R_{+} \\
      \alpha * Clip(\mathbf{X} - \beta, -1, 1),                      & \text{if } \mathbf{X} \in R
    \end{cases}.
    \label{eq:elastic_quantization}
\end{align}

Two-set elastic quantization provides a generic way to quantize any tensor, as shown in equation \eqref{eq:elastic_quantization}. In addition to quantization on model parameters to reduce the storage size, application on activations can also reduce floating operations further. For example, for low-parameter but high-computation operations such as multi-head attention, major computations happen between intermediate activations such as the query, key and values. Quantizing such activations can move significant amount of floating operations to quantized domains and improve computational efficiency, as will be discussed further in experiments.

\subsection{Squashed Weight Quantization (SqWQ)}
\label{subsec:squashed_weight_quantization}

Recently, squashed weight quantization (SqWQ) \cite{sqwq} was also proposed to reduce the quantization error. Squashed weight quantization aims to re-distribute the parameters into uniform distributions, as in equation \eqref{eq:sqwq_linear}, where $g$ is a gain factor in form of a vector.

\begin{align}
    \boldsymbol{y}&=Q(tanh(\mathbf{W}))*\boldsymbol{x}*e^{\boldsymbol{g}} + \mathbf{b}. \label{eq:sqwq_linear}
\end{align}

To achieve the re-distribution, the additional regularization loss $L_{Q}$ is added to the loss function as defined in equation \eqref{eq:sqwq_loss}, where $\lambda_{q}$ is the weight of regularization loss and $\sigma_{t}$ is the target standard deviation, both as hyper-parameters to be tuned.

\begin{align}
    L_{Q} = \lambda_{q} * ((stddev(\mathbf{W}) - \sigma_{t})^{2} + mean(\mathbf{W})^{2}). \label{eq:sqwq_loss}
\end{align}

By enforcing the parameters to be uniformly distributed, squashed weight quantization also preserves the utilization of the precious bits and demonstrated great performance for lower-bit models. 

\begin{table*}[!htp]
\centering
\scalebox{0.70}{
\begin{tabular}{cccccccccccccccc}
  \toprule
  \multirow{2}{*}{\textbf{Base Model}} & \multirow{2}{*}{\textbf{Quant}} & \multirow{2}{*}{\textbf{Precision}} & \multicolumn{9}{c}{\textbf{SUPERB Tasks}} & \multirow{2}{*}{\shortstack[c]{\textbf{Storage}\\\textbf{(MBs)}$\downarrow$}} & \multirow{2}{*}{\shortstack[c]{\textbf{FLOPs}\\\textbf{(Gs)}$\downarrow$}} & \multirow{2}{*}{\shortstack[c]{\textbf{QuantOPs}\\\textbf{(GBits)}$\downarrow$}} & \multirow{2}{*}{\shortstack[c]{\textbf{Runtime}\\\textbf{(Est. x)}$\downarrow$}} \\
 \cmidrule{4-12}
          & & & \textbf{ASR}$\downarrow$  & \textbf{KS}$\uparrow$ & \textbf{SF}$\uparrow$ & \textbf{PR}$\downarrow$ & \textbf{QbE}$\uparrow$ & \textbf{IC}$\uparrow$ & \textbf{ASV}$\downarrow$ & \textbf{SD}$\downarrow$ & \textbf{ER}$\uparrow$ & & & & \\
\hline
HuBERT (Base)\cite{hubert} & -- & fp16 & 6.42 & 96.59 & 0.88 & 5.41 & 7.36 & 97.15 & 5.11 & 6.20 & 64.92 & 189.14 & 153.14 & 0.00 & 1.38 \\
\hline
\multirow{13}{*}{\shortstack[c]{HuBERT\\(+FastConv\cite{fast_conv})}} & -- & fp16 & \textbf{7.06} & 96.62 & 0.89 & 6.05  & 6.91 & 97.28 & 5.30 & 6.32 & 65.00 & \textbf{184.42} & 110.79 & 0.00 & \textbf{1.00} \\
\cmidrule{2-16}
                          & \multirow{4}{*}{SqWQ\cite{sqwq}} & w8   & 9.69  & 96.88 & 0.88 & 7.30  & 6.19 & 96.65 &	5.88  & 6.52  & 62.83 & 99.65  & 82.24  & 1898.44  & \textbf{1.00} \\
                          &                                                 & w4   & 9.98  & 96.59 & 0.88 & 8.03 & 5.86 & 96.26 & 6.06 & 6.73 & 62.79 & 57.19  & 82.24	& 1054.69  & 0.89 \\
                          &                                                 & w2   & 12.56 & 94.22 & 0.86 & 11.79 & 5.27 & 94.02 & 6.31 & 7.12 & 62.38 & 35.95 & 82.24 & 632.81   & 0.83\\
                          &                                                 & w1   & 25.37 & 85.07 & 0.73 & 41.77 & 4.74 & 64.88 & 18.23 &	11.26 & 54.40 & 25.34  & 82.24	& 421.88   & 0.80\\
\cmidrule{2-16}
                          & \multirow{4}{*}{\shortstack[c]{BiT-L\cite{bit}\\(Linear\\Only)}} & w8a8 & 7.03  & 96.85 & 0.88 & 6.22 & 6.36 & 98.23 & 5.54 & 6.36 & 65.94 & 99.49 & 82.29 & 1898.44 & 1.00 \\
                          &                                                    & w4a4 & 8.58  & 96.56 & 0.88 & 7.15 & 6.40 & 96.10 & 5.55 & 6.26 & 64.12 & 57.02 & 82.29 & 527.34  & 0.81 \\
                          &                                                    & w2a2 & 10.80 & 95.88 & 0.86 & 8.79 & 5.62 & 97.47 & 5.68 & 6.55 & 63.49 & 35.79 & 82.29 & 158.20  & 0.76 \\ 
                          &                                                    & w1a1 & 12.23 & 94.94 & 0.86 & 10.49 & 5.99 & 96.49 & 6.55 & 6.87 & 63.06 & 25.17 & 82.29 & 52.73   & 0.75 \\ 
\cmidrule{2-16}
                          & \multirow{4}{*}{\shortstack[c]{BiT-LA\cite{bit}\\(Linear\\+Attention)}} & w8a8 & 7.07 & 97.21 & 0.89 & 6.30 & 6.40 & 98.10 & 5.56 & 6.24 & 65.77 & 99.54 & 11.82 & 3868.56 & 0.63 \\
                          &                                                    & w4a4 & 9.35  & 96.62 & 0.88 & 7.76 & 6.37 & 96.92 & 5.75 & 6.09 & 66.58 & 57.08 & 11.82 & 1074.60 & 0.25 \\
                          &                                                    & w2a2 & 12.68 & 95.07 & 0.85 & 12.56 & 5.23 & 95.02 & 7.40 & 6.94 & 63.00 & 35.84 & 11.82 & 322.38  & 0.15 \\ 
                          &                                                    & w1a1 & \textbf{15.96} & 93.83 & 0.78 & 22.96 & 5.63 & 93.01 & 6.83 & 7.62 & 61.68 & \textbf{25.23} & 11.82 & 107.46 & \textbf{0.12} \\ 
\hline
DistillHuBERT\cite{distillhubert} & -- & fp16 & 13.37 & 95.98 & 0.83 & 16.27 & 5.11 & 94.99 & 8.55 & 6.19 & 63.02 & 46.98 & 80.34 & 0.00 & 0.73 \\
\bottomrule
\end{tabular}
}
\caption{Evaluation of Quantization Techniques on SUPERB Tasks and Profiling Results.}\label{tab:superb}
\end{table*}

\section{Knowledge Distillation}
\label{sec:distillation}

During quantization-aware training, along with the quantization errors accumulated through operations, the gradient can also degrade through back-propagation \cite{bit,sqwq}. To mitigate the gradient degradation through operators, knowledge distillation \cite{qkd,distillation} has proven to be effective where the "student" model aims to imitate the outputs of the "teacher" model, regardless of the original loss function of the teacher model. There are different strategies to apply knowledge distillation for different scenarios and domains. Since quantization keeps the model architecture, meaning the student (quantized) model shares the same tensor shapes with the teacher (unquantized) model, we aim to distill not only the final output of the models but also the intermediate outputs and attention weights from each inner Transformer layers, as described in equation \eqref{eq:distill_final} and \eqref{eq:distill_layers}, where $MSE()$ is the mean square error operator, $\boldsymbol{y}_{T}$ and $\boldsymbol{y}_{S}$ are model outputs, $\boldsymbol{o}_{T, i}$ and $\boldsymbol{o}_{S, i}$ are intermediate outputs for layer $i$, $\boldsymbol{a}_{T, i}$ and $\boldsymbol{a}_{S, i}$ are attention weights for layer $i$ in teacher and student models.
\begin{align}
    L_{final} = MSE(\boldsymbol{y}_{T}, \boldsymbol{y}_{S}) + L_{layers}. \label{eq:distill_final} \\
    L_{layers} = {\textstyle\sum}_{i}(MSE(\boldsymbol{o}_{T, i}, \boldsymbol{o}_{S, i}) + MSE(\boldsymbol{a}_{T, i}, \boldsymbol{a}_{S, i})). \label{eq:distill_layers}
\end{align}

\section{Experiments}

\subsection{Experimental Setup}

We adopt the HuBERT\cite{hubert} (base) model as the baseline. For QAT, the same 960 hours of training set from LibriSpeech\cite{librispeech} for building the baseline model is used. The evaluation was performed on the 9 downstream tasks in the SUPERB\cite{superb} challenge including automatic speech recognition (ASR), keyword spotting (KS), slot filling (SF), phoneme recognition (PR), query by example (QbE), intent classification (IC), automatic speaker verification (ASV), speaker diarization (SD) and emotion recognition (ER). The tasks are labeled with up/down arrows showing the goals of the metrics as in Tables \ref{tab:superb} and \ref{tab:distill}. For example, ASR is measure in word error rates (WERs) therefore lower is better. In all tasks, the model serves as a speech representation extractor with parameters fixed after training. The implementation was built on top of fairseq\cite{fairseq} and torchaudio\cite{torchaudio}.

\begin{table*}[!htp]
\centering
\scalebox{0.82}{
\begin{tabular}{ccccccccccccc}\toprule
\multirow{2}{*}{\textbf{Base Model}} & \multirow{2}{*}{\textbf{Quant}} & \multirow{2}{*}{\textbf{Loss}} & \multirow{2}{*}{\textbf{Precision}} & \multicolumn{9}{c}{\textbf{SUPERB Tasks}} \\
\cmidrule{5-13}
          & & & & \textbf{ASR}$\downarrow$  & \textbf{KS}$\uparrow$ & \textbf{SF}$\uparrow$ & \textbf{PR}$\downarrow$ & \textbf{QbE}$\uparrow$ & \textbf{IC}$\uparrow$ & \textbf{ASV}$\downarrow$ & \textbf{SD}$\downarrow$ & \textbf{ER}$\uparrow$ \\
\hline
\multirow{9}{*}{\shortstack[c]{HuBERT\\(+FastConv\cite{fast_conv})}} & --  & --  & fp16 & 7.06  & 96.62 & 0.89 & 6.05  & 6.91 & 97.28 & 5.30  & 6.32  & 65.00 \\
\cmidrule{2-13}
  & \multirow{8}{*}{\shortstack[c]{BiT-LA\cite{bit}\\(Linear\\+Attention)}} & \multirow{4}{*}{HuBERT\cite{hubert}} & w8a8 & 8.32 & 93.34 & 0.89 & 7.30 & 6.31 & 90.26 & 6.51 & 7.25 & 63.16 \\
  & & & w4a4 & 10.59 & 92.25 & 0.85 & 8.68  & 5.89 & 90.82 & 6.57 & 7.14 & 64.22 \\
  & & & w2a2 & 14.29 & 91.72 & 0.81 & 14.46 & 4.80 & 88.68 & 8.71 & 7.85 & 60.29 \\
  & & & w1a1 & 18.93 & 91.07 & 0.77 & 25.92 & 5.22 & 83.80 & 8.05 &	8.73 & 58.50 \\
\cmidrule{3-13}
  & & \multirow{4}{*}{\shortstack[c]{Knowledge\\Distillation}} & w8a8 & 7.07  & 97.21 & 0.89 & 6.30  & 6.40 & 98.10 & 5.56 & 6.24 & 65.77 \\
  & & & w4a4 & 9.35  & 96.62 & 0.88 & 7.76  & 6.37 & 96.92 & 5.75 & 6.09 & 66.58 \\
  & & & w2a2 & 12.68 & 95.07 & 0.85 & 12.56 & 5.23 & 95.02 & 7.40 & 6.94 & 63.00 \\
  & & & w1a1 & 15.96 & 93.83 & 0.78 & 22.96 & 5.63 & 93.01 & 6.83 & 7.62 & 61.68 \\
\bottomrule
\end{tabular}
}
\caption{Comparison between Loss Functions for Quantized Model Training: HuBERT and Knowledge Distillation.}\label{tab:distill}
\end{table*}

For evaluating the resource consumption of the models, we extended the DeepSpeed \cite{deepspeed} tool to profile the models for 1) on-disk storage 2) floating point operations 3) quantization operations. The definition of these metrics are:

\begin{itemize}

\item \textbf{Storage}: The required space to store all model parameters, measured in megabytes (MBs). Parameters in fp16 are estimated to take 16 bits each, while quantized parameters take the same bits for annotated weight bits (e.g. 8 bits for w8).

\item \textbf{FLOPs}: The sum of floating operations (FLOPs) during the forward pass, measured in gigas (Gs).

\item \textbf{QuantOPs}: The sum of quantization (integer or binary) operations during the forward pass, measured in gigabits (GBits). QuantOPs are subjective to the number of bits in quantization. For example, for operations between a 8-bit integer and a 2-bit integer, a multiplication would take $2*8 = 16$ QuantOPs, while an addition would take $max(2, 8) = 8$ QuantOPs.

\item \textbf{Runtime}: The estimated runtime, measured in relative proportion (x) to the baseline HuBERT(+FastConv) model in fp16. Since FLOPs and QuantOPs are fundamentally different, their execution speeds are highly dependent on the hardware \cite{quant_speed}. In attempt to integrate the two types of operations in a hardware-agnostic fashion, we created an anchor point for conversion rate between FLOPs and QuantOPs. In Table \ref{tab:superb}, we assume the SqWQ-w8 model runs as fast as its fp16 counterpart and both have runtime 1.00, meaning $(110.79-82.24)=28.55$ Gs in FLOPs would run as fast as $1898.44$ GBits in QuantOPs. This conversion rate is applied to all models for runtime estimation in Table \ref{tab:superb}.

\end{itemize}

\subsection{Experimental Results}

\subsubsection{SUPERB Tasks and Profiling}
\label{subsubsec:superb}

Table \ref{tab:superb} presents the results for SUPERB tasks and model profiling. The HuBERT (Base) model \cite{hubert} as the first baseline has a relative high FLOPs given similar storage size compared with the alternative version (+FastConv) in which the convolutional feature extractor is replaced with a more efficient configuration from \cite{fast_conv} on top of \cite{wav2vec2}. Therefore, we applied quantization on top of the more efficient version HuBERT(+FastConv) and refer to it as the new baseline, which is 38\% more efficient in FLOPs. The precision for quantization in table \ref{tab:superb} are labeled with the number of bits for both model weights and activations. For example, w4a2 refers to 4 bits for model weights and 2 bits for activations for BiT. Since SqWQ focuses on quantizing the model weights, 8 bits were applied for activations across all setups \{w8, w4, w2, w1\}. 

Based on the HuBERT(+FastConv) model, 3 different quantization strategies were applied: 1) SqWQ 2) BiT (Linear Only) 3) BiT (Linear+Attention), for each different number of bits \{8, 4, 2, 1\} were all experimented. Since most SUPERB tasks demonstrate similar trends, we focus on the ASR task as the main metric.

To evaluate squashed weight quantization (SqWQ), results showed that the SqWQ-w8 model effectively reduced the storage size $(184.42 \rightarrow 99.65)$ with some degradation $(7.06 \rightarrow 9.69)$ on the ASR task. The SqWQ-w4 further reduced the storage $(99.65 \rightarrow 57.19)$ and also had lower QuantOPs $(1898.44 \rightarrow 1054.69)$ due to lower bits applied. The further degradation from SqWQ-w8 to SqWQ-w4 is milder $(9.69 \rightarrow 9.98)$ compared with fp16 to SqWQ-w8 $(7.06 \rightarrow 9.69)$, potentially benefiting from better utilization of bit by fitting the parameters into uniform distributions. The same trend applies to SqWQ-w2 and SqWQ-w1 models.

For models with BiT quantization applied to linear only (BiT-L), results showed similar trends as the SqWQ models. Both SqWQ and BiT-L applies to linear operations only and share similar trends in which the degradation increases and the number of bits becomes lower, with BiT-L showing smaller gaps from the fp16 baseline. For example, BiT-L-w1a1 is similar to SqWQ-w2 $(12.23 \leftrightarrow 12.56)$ and lower than SqWQ-w1 $(12.23 \leftrightarrow 25.37)$. In addition, since BiT applies quantization to activations as well, the models with lower bits also show lower complexity on QuantOPs, as observed between BiT-L-w1a1 and SqWQ-w1 $(52.73 \leftrightarrow 421.88)$. From the FLOPs and estimated runtimes, it is worth noting that although linear operations were moved to quantizated domains, a significant portion of FLOPs (82.24 out of 110.79) remained, which dominated the estimated runtime $(1.00 \rightarrow 0.75)$ even though the QuantOPs were reduced significantly $(1898.44 \rightarrow 52.73)$ comparing the BiT-L-w8a8 and BiT-L-w1a1 models. 

For models with BiT quantization applied to both linear and attention operations (BiT-LA), the storage did not reduce but slightly increased due to the scaling factors $\alpha$ and thresholds $\beta$ introduced in elastic quantization. For example, BiT-LA-w1a1 took more space than BiT-L-w1a1 $(25.23 \leftrightarrow 25.17)$ given similar bits. As quantization additionally applied to attention operations, the quantization error increased across all precisions, but tended to be worse for model with lower bits. For example, results showed that slight degradation from BiT-L-w8a8 to BiT-LA-w8a8 $(7.03 \rightarrow 7.07)$ but the gap was larger from BiT-L-w1a1 to Bit-LA-w1a1 $(12.23 \rightarrow 15.96)$. The major benefit from quantizing attention additionally is in computation. Comparing BiT-L-w1a1 and BiT-LA-w1a1, the FLOPs was significantly reduced $(82.29 \rightarrow 11.82)$ since the attention operators were moved to quantization domains. Although this also increased QuantOPs $(52.73 \rightarrow 107.46)$, it was insignificant for the low bits and the overall reduced estimated runtime $(0.75 \rightarrow 0.12)$.

In addition to quantization, other approaches such as neural architecture search are also of wide interest for efficient modeling. To compare the effectiveness of quantization, we included the results from DistillHuBERT \cite{distillhubert}, which also aims to improve the efficiency for HuBERT models. From the Table, comparable models included all three quantization strategies with 2 bits (SqWQ-w2, BiT-L-w2a2 and BiT-LA-w2a2), which were all around 35 MBs in storage. Results showed that all three quantization strategies provide lower word error rates $(12.56, 10.80, 12.68 \leftrightarrow 13.37)$ and lighter storage $(35.95, 35.79, 35.84 \leftrightarrow 46.98)$, with BiT-LA-w2a2 significantly outperformed on the FLOPs $(11.82 \leftrightarrow 80.34)$ and estimated runtime $(0.15 \leftrightarrow 0.73)$, which demonstrated that quantization is effective among efficient modeling techniques. 

\subsubsection{Knowledge Distillation for Quantized Model Training}

As described in section \ref{sec:distillation}, model training with quantization is challenging since quantization error accumulates through back-propagation. Therefore, we adopted knowledge distillation for minimizing not only the mean square error between model outputs but also intermediate layer outputs and attention weights. To investigate the impact from knowledge distillation, we trained another set of models with exactly the same quantization configurations, the same initialization from the baseline fp16 models, the same dataset but the loss function was changed to the original HuBERT loss \cite{hubert} for speech representation learning. These models were learning to predict masked frames instead of imitating the baseline fp16 model and there was no intermediate outputs involved. 

BiT quantization on both linear and attention operations (BiT-LA) was applied and the results were summarized in Table \ref{tab:distill}. We see that the trends are similar across precisions with models trained with HuBERT loss degrading more than models trained with knowledge distillation. For example, word error rates are $(18.93 \leftrightarrow 15.96)$ between HuBERT loss and knowledge distillation. The profiling metrics such as storage and runtime were omitted since they are independent from loss functions and can be found in Table \ref{tab:superb}. It is evident that knowledge distillation is effective for training quantized models by minimizing intermediate layer outputs to mitigate accumulated quantization errors through back-propagation.

\begin{figure}
    \centering
    \includegraphics[scale=0.36]{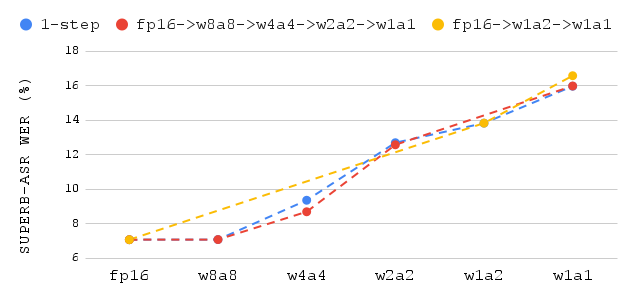}
    \caption{Comparison between One-step and Scheduled Quantization on SUPERB-ASR Task.}
    \label{fig:schedule}
\end{figure}

\subsubsection{One-Step vs. Scheduled Quantization}

In section \ref{subsubsec:superb}, the models were quantized directly from the original model in fp16 to the target precision. For example, BiT-LA-w1a1 was distilled directly from HuBERT(+FastConv)-fp16. This is referred to as "one-step" quantization, as opposed to the scheduled quantization suggested in \cite{bit}, where findings indicated that multi-step quantization involving intermediate precisions may improve the performance degradation. There are endless combinations to perform multi-step distillation, depending on the number of bits for weights and activations and how many bits to reduce at a time. In comparison, one-step quantization provides simpler setup but would potentially suffer from larger performance degradation. To understand the impact of scheduled quantization, two quantization schedules were experimented against the one-step results: 1) fp16 \textrightarrow w8a8 \textrightarrow w4a4 \textrightarrow w2a2 \textrightarrow w1a1 and 2) fp16 \textrightarrow w1a2 \textrightarrow w1a1 as suggested in \cite{bit}. The results were plotted in Figure \ref{fig:schedule} where the X-axis is for prevision and the Y-axis is the word error rate in the ASR task. Results showed that the models with scheduled quantization offered similar performance as their one-step counterparts, especially at the target precision w1a1. Given the many combinations of scheduled quantization, we draw the preliminary conclusion that no significant improvement was observed from scheduled quantization.

\section{Conclusion}
In this work, we investigated two novel quantization techniques: 1) robustly binarized Transformer \cite{bit} 2) squashed weight quantization \cite{sqwq}. The quantization was applied to HuBERT \cite{hubert} models for speech representation learning tasks. The experiments were evaluated on the SUPERB \cite{superb} benchmark and significant savings were observed both on storage and estimated runtime through quantization. For the aggressively binarized models, storage was saved by 86.32\% in megabytes (MBs) $(184.42 \rightarrow 25.23)$, estimated runtime was reduced by 88\% $(1.00 \rightarrow 0.12)$. For comparable configurations to DistillHuBERT \cite{distillhubert}, 2-bit models offered lower word rates $(12.68 \leftrightarrow 13.37)$ and estimate runtime $(0.15 \leftrightarrow 0.73)$ while still smaller in storage $(35.84 \leftrightarrow 46.98)$. With the growing sizes and computational complexity of modern models, we believe it is crucial to make models both compact and efficient so they are more accessible and deployable to resource-constrained environments such as edge devices. 

\section{Acknowledgement}
We would like to express our sincere gratitude for the following contributors: 1) Zechun Liu and Barlas Oguz from Meta AI for providing technical details and discussions on the "Robustly Binarized Transformer" work. 2) Xiaohui Zhang and Zhaoheng Ni from Meta AI for techincal discussions and implementation on top of fairseq and torchaudio. 3) Anuj Diwan from University Texas at Austin for technical discussions and optimization.

\bibliographystyle{IEEEbib}
\bibliography{strings,refs}

\begin{thebibliography}{10}

\bibitem{bit}
Zechun Liu, Barlas Oguz, Aasish Pappu, Lin Xiao, Scott Yih, Meng Li, Raghuraman
  Krishnamoorthi, and Yashar Mehdad,
\newblock ``Bit: Robustly binarized multi-distilled transformer,'' 2022.

\bibitem{sqwq}
Nikko Strom, Haidar Khan, and Wael Hamza,
\newblock ``{Squashed Weight Distribution for Low Bit Quantization of Deep
  Models},''
\newblock in {\em Proc. Interspeech 2022}, 2022, pp. 3953--3957.

\bibitem{hubert}
Wei{-}Ning Hsu, Benjamin Bolte, Yao{-}Hung~Hubert Tsai, Kushal Lakhotia, Ruslan
  Salakhutdinov, and Abdelrahman Mohamed,
\newblock ``Hubert: Self-supervised speech representation learning by masked
  prediction of hidden units,''
\newblock {\em CoRR}, vol. abs/2106.07447, 2021.

\bibitem{superb}
Shu{-}Wen Yang, Po{-}Han Chi, Yung{-}Sung Chuang, Cheng{-}I~Jeff Lai, Kushal
  Lakhotia, Yist~Y. Lin, Andy~T. Liu, Jiatong Shi, Xuankai Chang, Guan{-}Ting
  Lin, Tzu{-}Hsien Huang, Wei{-}Cheng Tseng, Ko{-}tik Lee, Da{-}Rong Liu, Zili
  Huang, Shuyan Dong, Shang{-}Wen Li, Shinji Watanabe, Abdelrahman Mohamed, and
  Hung{-}yi Lee,
\newblock ``{SUPERB:} speech processing universal performance benchmark,''
\newblock {\em CoRR}, vol. abs/2105.01051, 2021.

\bibitem{distillhubert}
Heng-Jui Chang, Shu-wen Yang, and Hung-yi Lee,
\newblock ``Distilhubert: Speech representation learning by layer-wise
  distillation of hidden-unit bert,''
\newblock in {\em ICASSP 2022 - 2022 IEEE International Conference on
  Acoustics, Speech and Signal Processing (ICASSP)}, 2022, pp. 7087--7091.

\bibitem{attention}
Ashish Vaswani, Noam Shazeer, Niki Parmar, Jakob Uszkoreit, Llion Jones,
  Aidan~N. Gomez, Lukasz Kaiser, and Illia Polosukhin,
\newblock ``Attention is all you need,''
\newblock {\em CoRR}, vol. abs/1706.03762, 2017.

\bibitem{gpt}
Tom~B. Brown, Benjamin Mann, Nick Ryder, Melanie Subbiah, Jared Kaplan,
  Prafulla Dhariwal, Arvind Neelakantan, Pranav Shyam, Girish Sastry, Amanda
  Askell, Sandhini Agarwal, Ariel Herbert{-}Voss, Gretchen Krueger, Tom
  Henighan, Rewon Child, Aditya Ramesh, Daniel~M. Ziegler, Jeffrey Wu, Clemens
  Winter, Christopher Hesse, Mark Chen, Eric Sigler, Mateusz Litwin, Scott
  Gray, Benjamin Chess, Jack Clark, Christopher Berner, Sam McCandlish, Alec
  Radford, Ilya Sutskever, and Dario Amodei,
\newblock ``Language models are few-shot learners,''
\newblock {\em CoRR}, vol. abs/2005.14165, 2020.

\bibitem{bert}
Jacob Devlin, Ming{-}Wei Chang, Kenton Lee, and Kristina Toutanova,
\newblock ``{BERT:} pre-training of deep bidirectional transformers for
  language understanding,''
\newblock {\em CoRR}, vol. abs/1810.04805, 2018.

\bibitem{roberta}
Yinhan Liu, Myle Ott, Naman Goyal, Jingfei Du, Mandar Joshi, Danqi Chen, Omer
  Levy, Mike Lewis, Luke Zettlemoyer, and Veselin Stoyanov,
\newblock ``Roberta: {A} robustly optimized {BERT} pretraining approach,''
\newblock {\em CoRR}, vol. abs/1907.11692, 2019.

\bibitem{xornet}
Mohammad Rastegari, Vicente Ordonez, Joseph Redmon, and Ali Farhadi,
\newblock ``Xnor-net: Imagenet classification using binary convolutional neural
  networks,''
\newblock {\em CoRR}, vol. abs/1603.05279, 2016.

\bibitem{quant_speed}
Tailin Liang, John Glossner, Lei Wang, and Shaobo Shi,
\newblock ``Pruning and quantization for deep neural network acceleration: {A}
  survey,''
\newblock {\em CoRR}, vol. abs/2101.09671, 2021.

\bibitem{quant_noise}
Angela Fan, Pierre Stock, Benjamin Graham, Edouard Grave, R{\'{e}}mi Gribonval,
  Herv{\'{e}} J{\'{e}}gou, and Armand Joulin,
\newblock ``Training with quantization noise for extreme model compression,''
\newblock {\em CoRR}, vol. abs/2004.07320, 2020.

\bibitem{quarl}
Srivatsan Krishnan, Max Lam, Sharad Chitlangia, Zishen Wan, Gabriel
  Barth-maron, Aleksandra Faust, and Vijay~Janapa Reddi,
\newblock ``Qua{RL}: Quantization for fast and environmentally sustainable
  reinforcement learning,''
\newblock {\em Transactions on Machine Learning Research}, 2022.

\bibitem{qat}
Benoit Jacob, Skirmantas Kligys, Bo~Chen, Menglong Zhu, Matthew Tang, Andrew~G.
  Howard, Hartwig Adam, and Dmitry Kalenichenko,
\newblock ``Quantization and training of neural networks for efficient
  integer-arithmetic-only inference,''
\newblock {\em CoRR}, vol. abs/1712.05877, 2017.

\bibitem{qkd}
Jangho Kim, Yash Bhalgat, Jinwon Lee, Chirag Patel, and Nojun Kwak,
\newblock ``{QKD:} quantization-aware knowledge distillation,''
\newblock {\em CoRR}, vol. abs/1911.12491, 2019.

\bibitem{ste}
Yoshua Bengio, Nicholas L{\'{e}}onard, and Aaron~C. Courville,
\newblock ``Estimating or propagating gradients through stochastic neurons for
  conditional computation,''
\newblock {\em CoRR}, vol. abs/1308.3432, 2013.

\bibitem{fast_conv}
Felix Wu, Kwangyoun Kim, Jing Pan, Kyu~J. Han, Kilian~Q. Weinberger, and Yoav
  Artzi,
\newblock ``Performance-efficiency trade-offs in unsupervised pre-training for
  speech recognition,''
\newblock {\em CoRR}, vol. abs/2109.06870, 2021.

\bibitem{distillation}
Geoffrey Hinton, Oriol Vinyals, and Jeff Dean,
\newblock ``Distilling the knowledge in a neural network,'' 2015.

\bibitem{librispeech}
Vassil Panayotov, Guoguo Chen, Daniel Povey, and Sanjeev Khudanpur,
\newblock ``Librispeech: An asr corpus based on public domain audio books,''
\newblock in {\em 2015 IEEE International Conference on Acoustics, Speech and
  Signal Processing (ICASSP)}, 2015, pp. 5206--5210.

\bibitem{fairseq}
Myle Ott, Sergey Edunov, Alexei Baevski, Angela Fan, Sam Gross, Nathan Ng,
  David Grangier, and Michael Auli,
\newblock ``fairseq: A fast, extensible toolkit for sequence modeling,''
\newblock in {\em Proceedings of NAACL-HLT 2019: Demonstrations}, 2019.

\bibitem{torchaudio}
Yao-Yuan Yang, Moto Hira, Zhaoheng Ni, Anjali Chourdia, Artyom Astafurov,
  Caroline Chen, Ching-Feng Yeh, Christian Puhrsch, David Pollack, Dmitriy
  Genzel, Donny Greenberg, Edward~Z. Yang, Jason Lian, Jay Mahadeokar, Jeff
  Hwang, Ji~Chen, Peter Goldsborough, Prabhat Roy, Sean Narenthiran, Shinji
  Watanabe, Soumith Chintala, Vincent Quenneville-Bélair, and Yangyang Shi,
\newblock ``Torchaudio: Building blocks for audio and speech processing,''
\newblock {\em arXiv preprint arXiv:2110.15018}, 2021.

\bibitem{deepspeed}
Jeff Rasley, Samyam Rajbhandari, Olatunji Ruwase, and Yuxiong He,
\newblock ``Deepspeed: System optimizations enable training deep learning
  models with over 100 billion parameters,''
\newblock in {\em Proceedings of the 26th ACM SIGKDD International Conference
  on Knowledge Discovery \& Data Mining}, New York, NY, USA, 2020, KDD '20, p.
  3505–3506, Association for Computing Machinery.

\bibitem{wav2vec2}
Alexei Baevski, Henry Zhou, Abdelrahman Mohamed, and Michael Auli,
\newblock ``wav2vec 2.0: {A} framework for self-supervised learning of speech
  representations,''
\newblock {\em CoRR}, vol. abs/2006.11477, 2020.

\end{thebibliography}

\end{document}